# Phonon interference effects in GaAs-GaP superlattice nanowires

*Chaitanya Arya[1], Johannes Trautvetter[1], Jose M. Sojo-Gordillo[1], Yashpreet Kaur[1], Valentina Zannier[2], Fabio Beltram[2], Tommaso Albrigi[3], Alicia Ruiz-Caridad[1], Lucia Sorba[2], Riccardo Rurali[3] and Ilaria Zardo[1]*

[1]Departement Physik, Universität Basel, 4056 Basel, Switzerland

[2]NEST, Istituto Nanoscienze-CNR and Scuola Normale Superiore, I-56127 Pisa, Italy

[3]Institut de Ciencia de Materials de Barcelona, ICMAB−CSIC, Campus UAB, 08193 Bellaterra, Spain

ABSTRACT. Fine-tuning the functional properties of nanomaterials is crucial for technological applications. Superlattices, characterized by periodic repetitions of two or more materials in different dimensions, have emerged as a promising area of investigation. We present a study of the phonon interference effect on thermal transport in GaAs-GaP superlattice nanowires with sharp interfaces between the GaAs and GaP layers, as confirmed by high-resolution transmission electron microscopy. We performed thermal conductivity measurements using the so-called thermal bridge method on superlattice nanowires with a period varying from 4.8 to 23.3 nm. The measurements showed a minimum of the thermal conductivity as a function of superlattice period up to room temperature, that we interpreted as an indication of the crossover from coherent to incoherent thermal transport. Notably, this effect is not destroyed by surface boundary or by phonon-phonon scattering, as the crossover trend is also observed at room temperature. Our results were corroborated by both *ab initio* lattice dynamics and semiclassical nonequilibrium molecular dynamics calculations. These findings provide insights into the wave-like and particle-like transport of phonons in superlattice nanowires and demonstrate the potential for engineering thermal properties through precise control of the superlattice structure.

INTRODUCTION

In the latest decades, continuous efforts have been carried out in order to enhance or suppress thermal transport in materials for numerous technological applications [1,2]. To this end, it is crucial to understand and control phonons through different methods. Namely, phonons are the quanta of lattice vibrations and are the main carriers of sound and heat in insulators and semiconductors. Specifically, low frequency (~kHz) phonons are responsible for sound transmission, while high frequency (~THz) phonons are responsible for heat transport. Therefore, one way to control and engineer heat transport consists in manipulating high frequency phonons by modifying the materials at the nanoscale, *i.e.* the length scale comparable to the phonon mean free path [3]. While a well-established route to achieve this goal is nano-structuring [4], more elaborated yet promising approaches consist in the combined use of nano-structuring and hetero-structuring to control the heat transport by means of interference effects that can be achieved in the coherent phonon transport regime [5–7].

Nanowires (NWs) are nanostructures characterized by a high aspect ratio, featuring a rod-like shape with diameters of the order of few to tens or hundreds of nanometers and lengths of the order of micrometers. They are promising candidates for studying phonons interference effects in different transport regimes because they offer unique possibilities, both in terms of geometry, *e.g.* with axial [8] and radial [9] heterostructures, as well as in terms of materials, as they release the strain in the radial direction, thus enabling the defect-free combination of lattice matched [10] or mismatched materials [11,12]. Furthermore, they allow the growth of high-quality junctions [13], and have characteristic length scales that nowadays can be controlled with high precision [14]. Particularly relevant for material engineering are superlattices (SLs), *i.e.* lattices made by different materials periodically alternated, that can be used to investigate the behavior of phonons scattered by interfaces. Typically, phonons scattered from single interfaces lose their phase information, leading to diffusive thermal transport, whereas a periodic repetition of interfaces can lead to constructive interference, resulting in coherent phonon transport [15–18]. For these phenomena to occur, it is crucial for the interfaces to be as clean as possible, *i.e.* defect-free and sharp. On perfectly smooth interfaces, phonons scatter specularly and can interfere constructively with the reflected phonons provided that they are in phase, resulting in altered dispersion relations and in the formation of bandgaps [2]. The presence of periodically repeated interfaces can further modify the vibrational properties or phonon spectra, as wave interferences can influence the density of states and group velocities of the phonons [16].

While there is a significant potential for wave interference effects to impact thermal devices, demonstrations of these effects on macroscopic thermal transport quantities are still discussed. SLs provide an ideal platform for studying and understanding coherent phonon effects on macroscopic thermal properties. Depending on the period of the SL and on its relation to the coherence length of phonons, a wave-particle crossover is expected to occur [19,20]: when the SL period is smaller than the coherence length of the phonons, the wave nature of phonons becomes evident, leading to the appearance of interference effects; on the other hand, when the SL period is larger than the phonon coherence length, phonons are better described as particles that undergo individual and uncorrelated scattering events. The experimental observation of coherent heat conduction has been first reported in 2012 by Luckyanova et al. [20]. In that study, the thermal conductivity ($\kappa$) of

GaAs/AlAs SLs with a varying number of periods was measured using time-domain thermal reflectance technique in the temperature range from 30 to 300 K. In the coherent regime, the phonon phase information is preserved at the interfaces of the SL. The superposition of Bloch waves leads to the creation of stop bands, effectively modifying the phononic band structure. Consequently, they observed a linear dependence of the thermal conductivity on the total SL thickness over a temperature range from 30 to 150 K, suggesting that phonons can maintain phase coherence across multiple interfaces. On the other hand, in the incoherent regime, phonons are diffusively scattered at each internal interface, causing them to lose their phase information. The interfaces act as independent thermal resistors, leading to an effective thermal conductivity perpendicular to the interfaces that is approximately independent of the number of layers of the SL and that tends to the alloy of the two constituent materials [21].

Another significant signature of thermal transport across SLs is the presence of a minimum in κ as the interface density (number of interfaces for unit length) varies [20,23,24]. This minimum is an indication of a transition from particle-like to wave-like transport of phonons, as first proposed by Simkin and Mahan [20]. The first conclusive experimental evidence of this transition from particle-like (incoherent) to wave-like (coherent) processes was obtained through measurements of lattice thermal conductivity as a function of interface density in epitaxial oxide SLs [21]. This wave-particle crossover manifests as the existence of a minimum in κ as a function of interface density [19].

However, the transition from coherent to incoherent behavior has mainly been explored in 2D superlattices and, similar studies in 1D nanostructures are less prevalent. One of the open questions in these systems is the impact of surface boundary scattering on the coherence of phonons and their interference. Yet, 1D systems offer greater potential for heterostructuring and the creation of high-quality NW junctions. We have previously demonstrated the tunability of the phononic spectrum by analyzing the dependence of both acoustic and optical phonon modes on the SL period. As the SL period increases, the number of phonon modes also increase, which can be attributed to the larger number of atoms per unit cell [26]. In this work, we investigated the phonon interference effects in GaAs-GaP SL NWs by measuring the thermal conductivity using the thermal bridge method.

MATERIALS AND EXPERIMENTAL METHODS

GaAs-GaP SL NWs with different periodicities and uniform thickness used for this study were grown using Au-assisted chemical beam epitaxy (CBE) on a GaAs (111)B substrate with the same conditions detailed in prior work [27]. The NWs are composed of four segments, as depicted in the schematic in Figure 1(a). The bottom segment consists of 0.5 μm long GaAs, followed by a GaP stem of about 1 μm, an alternating GaAs/GaP superlattice segment, and a GaP segment of about 1 μm at the top. Figure 1(b,c) shows the high resolution transmission electron microscopy (TEM) image of two exemplary SL segments. The contrast shows the sharp interfaces between GaAs and GaP layers for a 4.8 nm and a 10 nm SL period NW. The core diameter of superlattice NW ranges from 30 nm to 50 nm and the shell around the NW has 20 nm thickness and it mainly consists of GaP. We performed thermal transport experiments on SL with periodicities ranging from 4.8 nm to 23.3 nm. Most of the SL were composed of 100 repetitions, while samples with 14.6 nm and 23.3 nm period have 30 repetitions without a top GaP segment. The SL samples investigated in this work are listed in the Table 1 of the Supporting Information S1 along with their description.

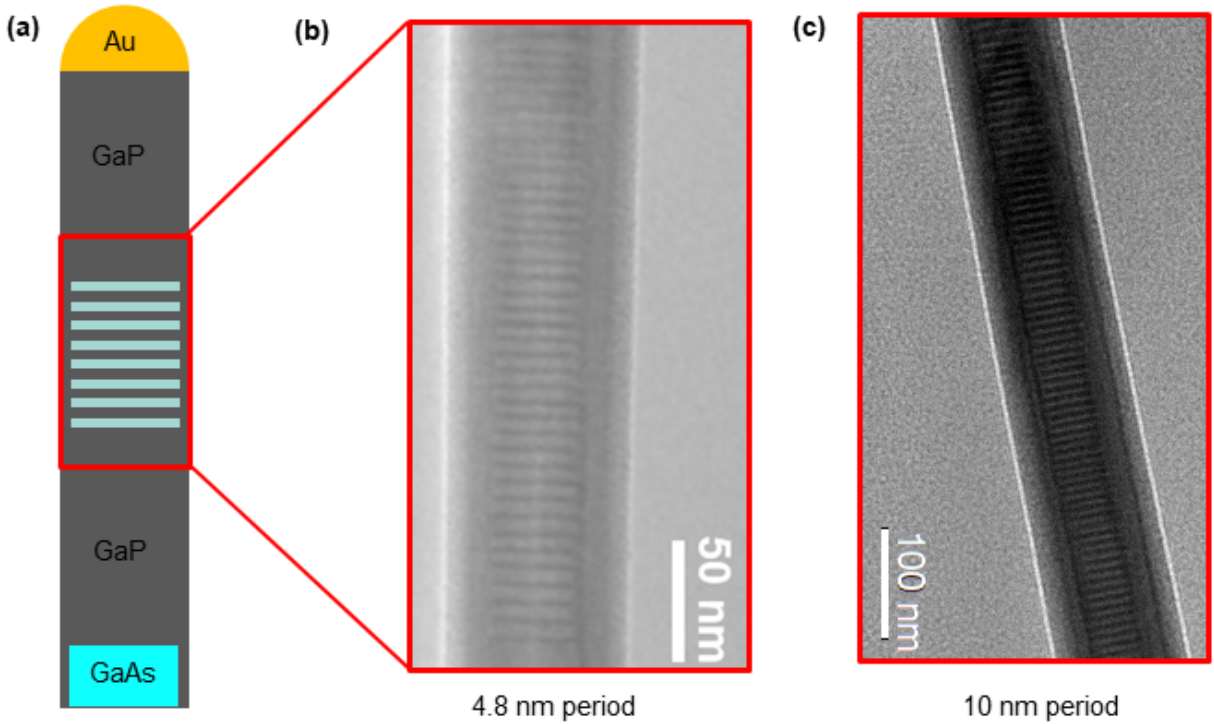

**Figure 1.** Superlattice nanowire sample. (a) Schematic of a SL NW composed of a GaAs/GaP base segment, a central SL segment and a top GaP segment. (b) TEM image of the SL segment in the middle of the NW for 4.8 nm period sample (c) and for 10 nm period sample.

The thermal conductivity of SL NWs is measured using the well-established suspended thermal bridge method proposed by Shi et al. [28]. To ensure thermal isolation, we fabricated a microdevice consisting of gold deposited on 0.5 mm long suspended SiNx beams. These SiNx beams supported two suspended platforms in the center, which contained platinum resistors acting as a heater or temperature sensor, while the beams have gold patterned lines which lead to contact pads for electrical connections, as shown in Figure 2(a,b). In the device fabrication process, the design of gold lines and platinum resistors on SiNx was accomplished using optical and electron beam lithography techniques respectively, while gold and platinum metals were deposited using electron beam evaporation followed by lift-off in acetone. Subsequently, the device was prepared for suspension. This involved a series of etching steps, both dry and wet, to create the required suspended structure by selectively removing the materials. Finally, the SiNx membranes were carefully cut to create a gap between the two platforms. This was done using focused ion beam (FIB) to ensure a high control in the gap dimensions (~1 µm). This is crucial the SL NWs used in our study are not very long and require small gaps between bridges. The Supporting Information S2 provides more details on the specific steps and parameters involved in the device fabrication process.

The suspended devices were calibrated in a probe station (Janis ST-500) in vacuum (10-5 - 10-6 mbar). Experiments at low temperature (below RT) were performed with liquid helium cooling. Electrical measurement of the resistors featured by the suspended devices was carried out using a source-meter unit (Keithley 4200A-SCS Parameter Analyzer) in a four-wire configuration. Contact with the chip was achieved using multiprobe tips.

Prior to the thermal conductivity measurements, these resistors required a calibration. During this process, the change in resistance of the platinum resistors is measured with respect to the base temperature variations and with respect to the power dissipated in each meander. This data allows for the calculation of the coefficient of thermal resistance $dR/dT$ as well as the beam conductance $G_B = (dR/dP)/(dR/dT) = d\Delta T/dP$ of both platforms. Both parameters are required for subsequent measurements (more details in Supporting Information S3). After the calibration process, GaAs-GaP NWs are transferred from the original substrate with vertical arrays of NWs onto the suspended devices between the two platforms using a hydraulically actuated micromanipulator. To measure the thermal conductance of the NW, a controlled temperature difference is created between the two platforms. One platform is heated while the change in temperature on the second platform, due to thermal transport through the NW, is measured as a function of the applied heating power. The temperature of the platforms is obtained by measuring the resistances using a four-point probe technique, which serves as a reliable indicator of temperature variations thanks to the linear temperature coefficient of resistance of the platinum lines of the resistor. The heat flux across the NW is then calculated by measuring the power dissipated in the resistors. As depicted in Figure 2(c), the NW bridges both platforms, forming a thermal pathway. When a bias current is applied through the heater platform, heat is generated, and some of this heat is transferred through the NW to the sensor platform. As a result, the temperature of the sensing resistor increases, and thus, its resistance. This parameter is then simultaneously recorded as a function of the applied power at the heater side. Hence, the NW thermal conductance is calculated as (details in Supporting Information S4) [28,29]:

$$G_N = G_{B,S}\frac{d\Delta T_S}{dP}\left(\frac{G_{B,H}}{G_{B,S}}\cdot\frac{d\Delta T_H}{dP} - \frac{d\Delta T_S}{dP}\right)^{-1}$$

where $\frac{d\Delta T_H}{dP}$ and $\frac{d\Delta T_S}{dP}$ are the heater and sensor temperature increase rates, respectively, as a function of the dissipated heater power, and $G_{B,H}$ and $G_{B,H}$ are the heater and sensor beam conductances, respectively, calculated as $\frac{d\Delta T_H}{dP}$ when the device is measured without a NW sample bridging both platforms. Finally, the thermal conductivity of the NW is calculated as $\kappa = \frac{4G_N L}{\pi D^2}$, where L is the NW suspended length and D is the diameter of the wire.

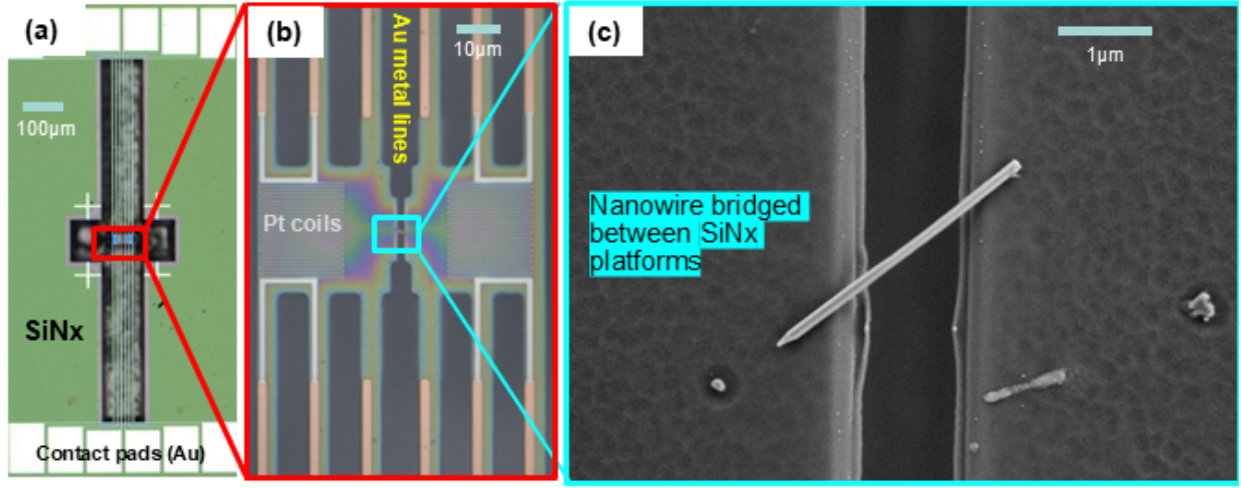

**Figure 2.** Suspended microdevice for measuring the thermal conductivity of NWs. (a) Optical image of the suspended device. (b) Left and right Pt resistors at the center are supported by long SiNx beam with gold metal lines on top. (c) SEM image of a SL NW suspended between the SiNx platforms.

RESULTS AND DISCUSSION

We used the suspended thermal bridge method to measure the thermal conductivity ($\kappa$) of GaAs-GaP SL NWs for various superlattice periods at bath temperatures from 16 K to 350 K. The NWs SL period ranged from 4.8 nm to 23.3 nm. The temperature dependent $\kappa$ results are summarized in Figure 3. Figure 3 (a) shows the measured $\kappa$ of four NWs with the smallest SL period (4.8 nm) at 300 K, depicting the typical statistical distribution and reproducibility in $\kappa$ for a given period. The slight variation in conductivity of different NWs can be attributed to the different surface roughness, variation in the diameter ranging from 114 nm to 130 nm (Supporting Information S1), and contact thermal resistances with the platforms.

Further, we perform a temperature dependent measurement of κ for the 4.8 nm and 10 nm period SL NWs, as well as for a GaP reference NW from 16 K to 350 K, shown in Figure 3 (b). At each bath temperature, experiments were performed for both directions of thermal bias, and the value of the κ was calculated averaging the results of the two measurements. The temperature dependence found is characteristic of NWs [30] and resemble closely that of Si NWs with diameters between 40 and 120 nm as reported by Li and coworkers [31]. Likewise, in our studied NWs, the maxima in κ are shifted to higher temperatures as the additional contribution of boundary scattering masks Umklapp scattering. Moreover, the κ of SL NWs peaks at higher temperatures - 173 K for the 10 nm period and 218 K for the 4.8 nm period - compared to the GaP reference nanowire, which peaks at 161 K. This is markedly different from the case of bulk GaAs, where κ exhibits a sharp maximum around 20K followed by a characteristic Umklapp scattering-driven $T^{-1}$ dependence [32,33]. A similar behavior is found in bulk GaP with κ peaking at 30 K [34]. Since our three studied NWs share comparable diameter, we expect the similar level of size confinement. Therefore, we attribute the observed shift in maxima of κ to higher temperatures to the presence of periodic interfaces in the SL structures, which would further suppress high-frequency phonons beyond the effects of boundary scattering.

Figure 3 (c) shows the low temperature range of the measured κ on logarithmic scale from 15 to 70 K. We roughly see a $T^{1/2}$ dependency for the three studied sample NWs. This trend is remarkably different *e.g.* from the case of bulk GaAs, where κ exhibits a $T^3$ dependence [32,33]. This suggests that, at these low temperatures, the boundary scattering produced by the finite diameter of the NWs dominates the phonon propagation as opposed to the contribution of the SL, which would require phonon coherency lengths (roughly the same order of magnitude as the phonon mean free path) proportional to the SL periods to have a significant impact. Therefore, since we do not observe a clear relationship between the trends in κ of pure GaP NW and the SL NWs studied, we cannot conclude that the effect of the SL is significant for T < 70 K.

In order to further understand these measurements, we plotted in Figure 3 (d) the computed κ of 5.1 nm and 10.2 nm GaAs/GaP SL bulk material and, additionally, the κ of a pure GaP NW with a diameter of 76 nm. There results were based on *ab initio* calculations, where we used the VASP code [35] to perform density-functional theory (DFT) calculations of the harmonic and anharmonic force constants and then solve the Boltzmann Transport Equation (BTE). In the case of SLs, we used the method implemented in almaBTE [36] that allows bypassing the explicit calculation of the phonon scattering rates in the superlattice unit cell —which would be unfeasible at the *ab initio* level— and rather relies on the phonon properties of the constituent materials, which were carefully determined in advance [37]. The reliability of this methodology is witnessed by the very good agreement with experimental results on Si/Ge [38] and InAs/GaAs SLs [39]. As coherence is destroyed outside the SL computational cell used for the solution of the BTE, in order to avoid artifacts, we use a supercell made of 60 repetitions of the wurtzite (WZ) unit cell for all periods (convergence tests with 120 repetitions were satisfactorily conducted). This length allows to accommodate almost exactly all the periods displayed in the plot (full details on the calculations are provided in the Supporting Information S5). Here, we notice that the absolute values obtained for κ are higher, as additional effects such as sample-device contact resistances or boundary scattering in the case of the SLs could not be taken into account. Nevertheless, in all cases, we also observe a shift of the maxima of the thermal conductivity from bulk values (12 K for GaP) to higher temperatures (125 K for 5.1 nm and 115 K for 10.2 nm SL period), in accordance with the experimental observations. We also notice how the SL structure itself seems also to play a role in shifting the temperature at which κ reaches a maximum, particularly for short periods and around the coherent-incoherent crossover, which occurs for SL periods of ca. 8 nm as exposed by SL period dependent thermal conductivity measurements (see Supporting Information Figure 5, where we plot the theoretical $\kappa(T)$ for different periods of a bulk SL).

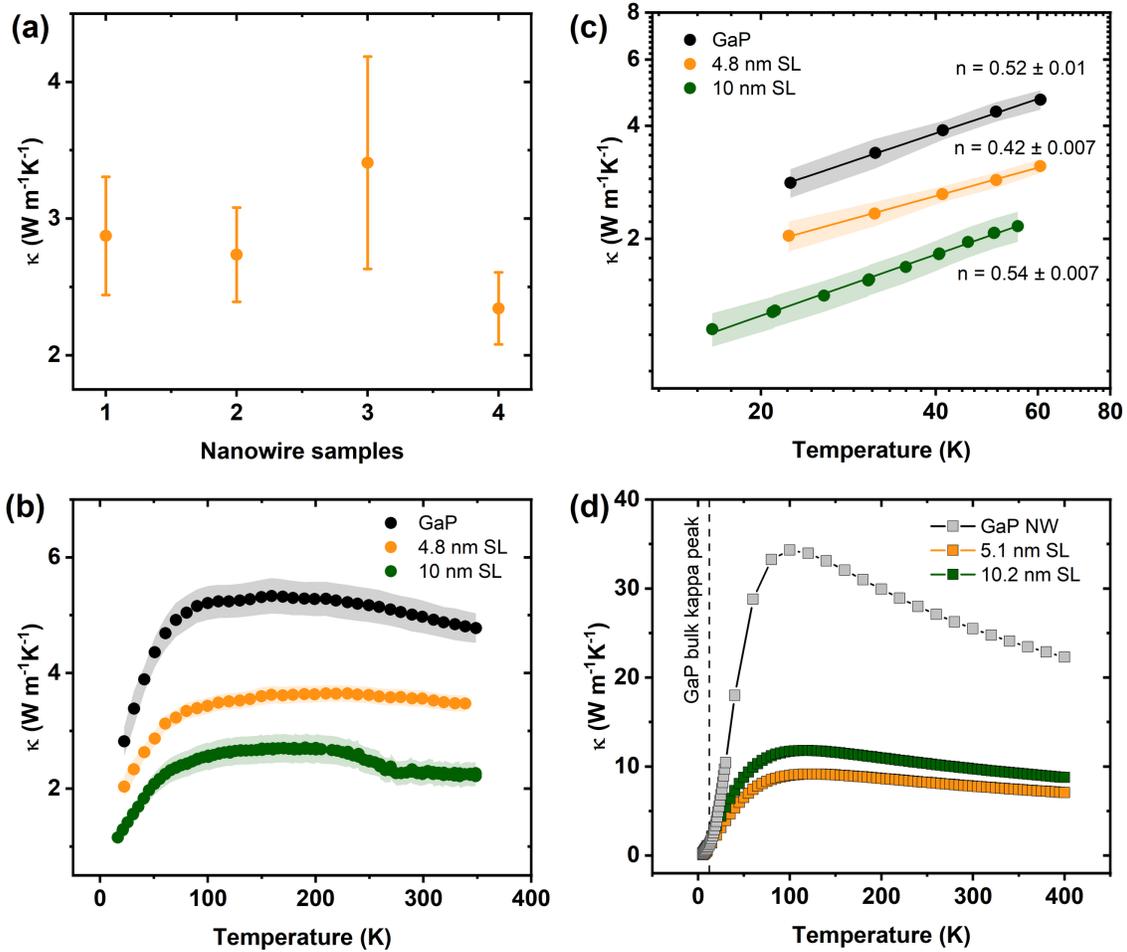

**Figure 3.** (a) Thermal conductivity of four different NWs featuring a 4.8 nm superlattice period. (b) Thermal conductivity measured for NWs featuring 4.8 nm (red) and 10 nm (blue) superlattice period, and GaP reference NW (black) from 16 K to 350 K. The error is represented by the shaded area. (c) Low temperature experimental data on a logarithmic scale. The solid lines are linear fits to the logarithmic data. The error is represented by the shaded area. (d) Computed thermal

conductivity of 5.1 nm and 10.2 nm superlattice period bulk material and a GaP NW with a diameter of 76 nm. The vertical dashed line shows the temperature at which the thermal conductivity of bulk GaP is maximum, *i.e.* 12 K. In all plots, circle symbols represent experimental data, while square ones stand for theoretically calculated values.

Subsequently, we systematically measured κ as a function of superlattice periods at 300 K and 140 K, see Figure 4 (a). With the exception of the longest period (23.3 nm), for each SL period, we have measured at least two wires under the two directions of thermal bias, and on each wire the measurements were repeated 5 times to verify the reproducibility of the measurements. In Figure 4 (a), we plot the average of the measured thermal conductivities for each period as a function of the SL period. Noteworthy, the measured thermal conductivity is the result of the combined contributions of the SL in series with the one of the pure GaP segments (see Figure 1) and of the contribution of contact resistances.

Starting from large periods, with decreasing SL period length, the thermal conductivity value decreases. This decrease in κ can be attributed to an increase in phonon scattering due to the increased number of interfaces. For this range of SL periods (>8 nm), the phonon coherence length is smaller than the period length, thus, increasing the periodicity of the lattice simply creates further number of interfaces (per unit length) causing higher phonon scattering, and thus yielding a lower overall κ. Around SL period sizes of ~8 nm, the conductivity value reaches a minimum, and then it starts to increase again with decreasing period length. This occurrence of a minimum in the thermal conductivity for SL NWs as a function of decreasing SL period is a signature of the transition from an incoherent to coherent phonon transport regime. In the coherent regime, *i.e.* for SL periods smaller than the coherence length, phonon phase information is preserved at the interfaces of the SLs and interference between phonons and their reflections occurs, leading to a modification of the phonon dispersion with the formation of mini-bands and possibly with the opening of forbidden energy gaps. In particular, the number of interfaces determines the number of mini-bands formed, which is directly related to the average phonon group velocity. Therefore, an increase in the number of interfaces yields an enhancement of the thermal conductivity, as observed for the NWs with a period of 4.8 nm in Figure 4 (a). Comparing the data set obtained for the two base temperatures, the trend is similar, though the minimum of κ is more pronounced at low temperature, and the minimum seems to shift towards longer SL periods, as expected from the increase in phonons mean free path (MFP) at low temperature and in agreement with previous finding [17].

Our results are particularly promising as they demonstrate coherent phonon interference effect in 1D systems, despite their high surface-to-volume ratio. Namely, in order to maintain the phase coherence, phonons must scatter specularly at the surface boundaries of nanostructures. Typically, the phonon-boundary scattering is diffusive and suppresses the phonon MFP. Our results, instead, indicate the preservation of phonon coherence up to room temperature. We could speculate that the presence of the shell around the SL depicted in Figure 1 (b,c) fosters specular scattering and thus helps preserving coherency. In particular, it is worth noticing that the GaP shell around the SL provides a space layer between the SL itself and the oxidized GaP out layer. The SL/GaP interface has a better quality than the one between GaP and amorphous oxide, and could, therefore, provide higher rate of specular scattering.

Furthermore, these results shine a light on the possibility to obtain coherent phonon in highly strained systems. Indeed, GaAs and GaP have a 3.7% lattice mismatch. The ability to achieve coherent phonon transport in such nanostructures highlights the potential for engineering thermal properties at the nanoscale, even in systems where strain effects are more pronounced.

Subsequently, we corroborated again our experimental results against two different computational frameworks. Firstly, in Figure 4 (b) we report our results of κ as a function of the period of a bulk GaAs/GaP SL based again on DFT calculations as previously described. As it can be seen, we obtain a minimum in κ(L) around 5 nm at both the temperatures considered. Similar to the experiments, for small period we observe a sharp increase in κ, the fingerprint that the transport regime has become coherent and interfaces does not act anymore as individual, independent barriers. The increase of the thermal conductivity on the diffusive regime depends on the atomistic structure of the interfaces and thus on the associated thermal boundary resistance (TBR). In the Supporting Information S5 we also show data for idealized, atomically flat interfaces, where phonon scattering is entirely dominated by Umklapp processes, due to the low value of the TBR, and κ(L) for large L's is flat (the contribution of the TBR of each interface is so small that the difference between having *e.g.* 4 or 5 interfaces is negligible).

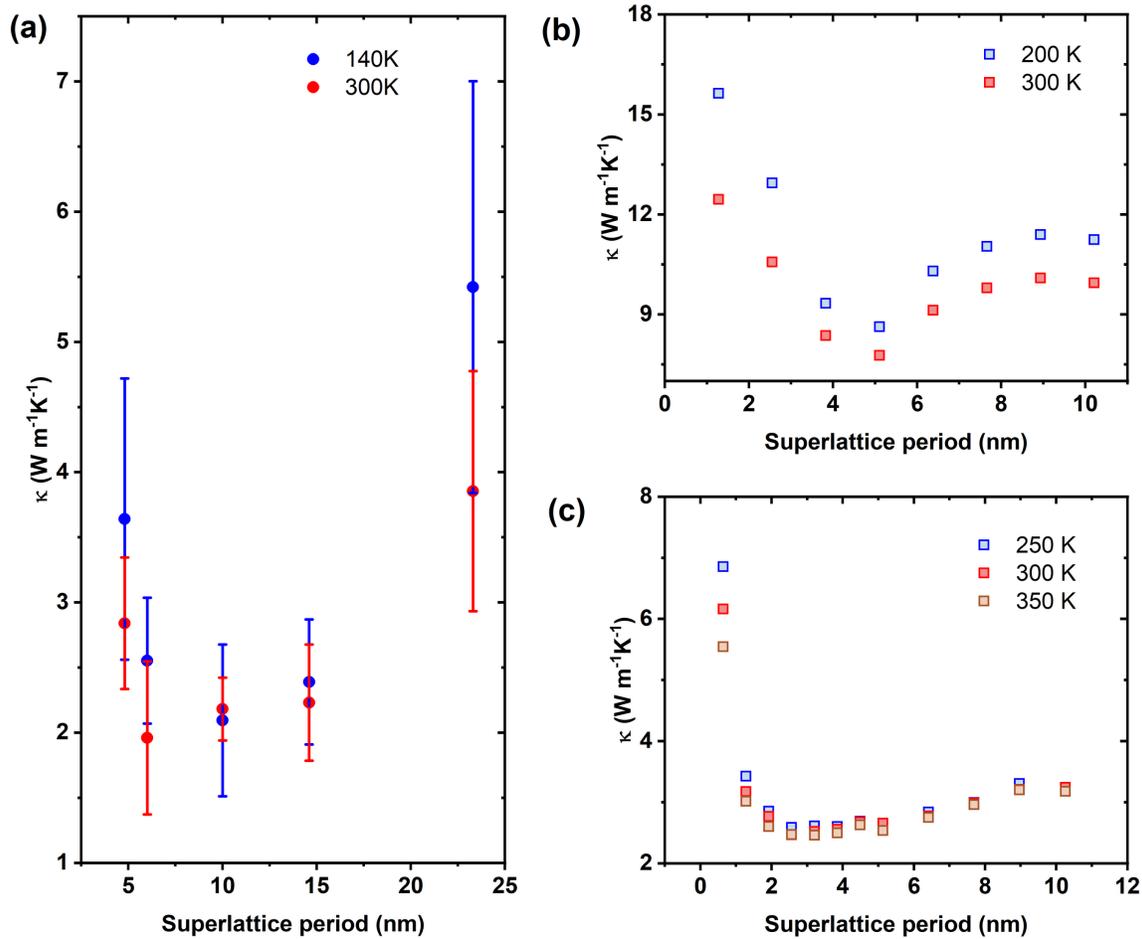

**Figure 4.** (a) Average thermal conductivity of superlattice nanowires as a function of superlattice period at 300 K (red) and at 140 K (blue). The errors correspond to the standard deviations of multiple measurements. (b) Computed thermal conductivity as function of the period in GaAs/GaP SLs obtained from DFT/BTE and (c) NEMD calculations. The DFT/BTE calculations are carried out in a bulk system, while in NEMD we considered a NW of 5 nm diameter. In all plots, circle symbols represent experimental data, while square ones stand for theoretically calculated values.

Secondly, we have also performed computational experiments based on nonequilibrium molecular dynamics (NEMD) with the LAMMPS code [40] and a bond-order potential [41] (full details on the calculations are provided in the Supporting Information S5). These calculations are complementary to the *ab initio* ones discussed above. While they cannot be expected to have a predictive power, being based on an empirical interatomic potential not especially designed to reproduce thermal transport properties, they allow to account explicitly for the NW geometry and do not need to rely on any phenomenological model for the TBR. Our results for a GaAs/GaP NW SL with a diameter of 5 nm and a total length of 100 nm and for two different thermal bias conditions are shown in Figure 4 (c), where we plot the thermal conductivity as a function of the SL period. We find a minimum in κ which falls at shorter period length with respect to *ab initio* results. This is a qualitative indication that the NW geometry and a more realistic description of the TBR do not hinder per se the appearance of a minimum in κ(L) and thus the onset of a coherent transport regime at short periods.

CONCLUSIONS

We performed experiments and numerical simulations to investigate the thermal conductivity of GaAs-GaP SL NWs with varying superlattice periods, exploring the transition from incoherent to coherent phonon transport as the superlattice period decreases. Experimental results show the temperature dependence of the highly suppressed thermal conductivity of SL nanowires due to the combined effect of boundary and TBR scatterings. More remarkably, we observed a decrease in thermal conductivity with decreasing period length, reaching a minimum at around 8 nm, indicative of an increased phonon scattering at interfaces that later crossovers towards a coherent transport for shorter SL periods, where the thermal conductivity was found to increase again. *Ab*

*initio* calculations support these findings, showing a minimum thermal conductivity at approximately 5 nm. Computational experiments using nonequilibrium molecular dynamics, where the nanowire geometry is explicitly accounted for, vouch for the generality of these observations. We have demonstrated the preservation of coherence of phonons up to room temperature in NWs, despite the importance of surfaces in these nanostructures, possibly arising from specular boundary scattering. Our findings display an interesting way to tune thermal properties by carefully designing a material system such as a superlattice nanowire.

AUTHOR INFORMATION

**Corresponding Author**

E-mail: Ilaria.zardo@unibas.ch

**Author Contributions**

I.Z. conceived the experiment and supervised the project. C.A., J.T., and J.M.S.-G. performed the thermal conductivity measurements, which were analyzed by C.A. and J.M.S.-G. C.A. and Y.K. designed and fabricated the thermal bridge devices. The samples were grown by V.Z., F.B., and L.S, while C.A. and J.T. transferred the wires onto the thermal bridge devices. T.A. and R.R. performed the theoretical calculations. A.R.C. performed EDX analysis and TEM investigations. C.A., J.M.S.-G., R.R., and I.Z. wrote the manuscript with contributions of all authors. All authors have given approval to the final version of the manuscript.

**Funding Sources**

Swiss National Science Foundation grant (Grant No. 200021_184942), Marie Sklodowska-Curie QUSTEC (grant agreement no. 847471), European Research Council (grant agreement No 756365).

ACKNOWLEDGMENT

We thank Arianna Nigro for fruitful discussion and technical support. R.R. thanks Jesús Carrete for useful discussions. This project has received funding from the Swiss National Science Foundation grant (Grant No. 200021_184942), from Eucor, The European Campus (Marie Sklodowska-Curie QUSTEC grant agreement no. 847471), and from the European Research Council (ERC) under the European Union's Horizon 2020 research and innovation program (grant agreement No 756365). R.R. acknowledges financial support by MCIN/AEI/10.13039/501100011033 under grant PID2020-119777GB-I00, and the Severo Ochoa Centres of Excellence Program under grant CEX2023-001263-S, and by the Generalitat de Catalunya under grant 2021 SGR 01519. We thank the Centro de Supercomputación de Galicia (CESGA) for the use of their computational resources. V.Z acknowledges financial support from the PRIN project 20223WZ245 – GROUNDS – "Growth and optical studies of tunable quantum dots and superlattices in semiconductor nanowires".

**Conflict of Interest Statement**

The authors declare no conflict of interest.

**Data Availability Statement**

The data that support the findings of this study are openly available in ZENODO at ##, reference number ##.

# Phonon interference effects in GaAs-GaP superlattice nanowires

*Chaitanya Arya[1], Johannes Trautvetter[1], Jose M. Sojo-Gordillo[1], Yashpreet Kaur[1], Valentina Zannier[2], Fabio Beltram[2], Tommaso Albrigi[3], Alicia Ruiz-Caridad[1], Lucia Sorba[2], Riccardo Rurali[3] and Ilaria Zardo[1\**

[1]Departement Physik, Universität Basel, 4056 Basel, Switzerland

[2]NEST, Istituto Nanoscienze-CNR and Scuola Normale Superiore, I-56127 Pisa, Italy

[3]Institut de Ciencia de Materials de Barcelona, ICMAB−CSIC, Campus UAB, 08193 Bellaterra, Spain

E-mail: Ilaria.zardo@unibas.ch

## S1: Superlattice nanowire samples

Table 1: Summary table of all samples used in this study.

| GaAs thickness (nm) | GaP thickness (nm) | SL period (nm) | Repetition | SL segment length (μm) | Nanowire diameter (nm) | SL position |
|---|---|---|---|---|---|---|
| 2.0 | 2.8 | 4.8 | 100 | 0.48 | 114 - 130 | center |
| 3 | 3 | 6 | 100 | 0.6 | 120 - 140 | center |
| 4.2 | 5.8 | 10 | 100 | 1 | 130 | center |
| 5 | 5 | 10 | 100 | 1 | 110 - 130 | center |
| 6.5 | 8.1 | 14.6 | 30 | 0.438 | 100 | tip |
| 10.7 | 12.6 | 23.3 | 30 | 0.699 | 90 | tip |

## S2: Device fabrication process

To accurately measure the thermal conductivity of a single nanowire (NW), a sensitive platform is required to measure the temperature gradient across the NW and calculate the heat flux through it. One crucial requirement is that the platform must be isolated from the surrounding environment to prevent heat dissipation. Therefore, the experiments are typically conducted in a vacuum environment to minimize heat loss to the surroundings. In order to calculate the NW's thermal conductivity, one end of the wire is heated while the temperature at this end is precisely controlled and measured by recording the change in resistance. Simultaneously, the temperature at the other NW's end is also measured. Fourier's law is then used to compute the thermal conductivity of the NW. By knowing the heat flux and the temperature gradient across the NW, the thermal conductivity can be determined [1].

Figure S1 shows a scanning electron microscope (SEM) image of the suspended device fabricated on a silicon substrate. The device consists of two suspended platforms made of silicon nitride ($SiN_x$). These platforms are equipped with metal lines and are separated by a small gap. The central region of the device consists of platinum resistors, while gold is deposited on the lines. The platinum resistors have a higher resistance compared to the gold lines, resulting in the majority of the power being dissipated in the resistors. The gold pads, which are not suspended, serve as heat sinks. To perform the electrical measurements, multiprobe tips are employed to contact the gold pads. These tips enable the application of current and the measurement of voltage, allowing for precise control and monitoring of the electrical properties of the device.

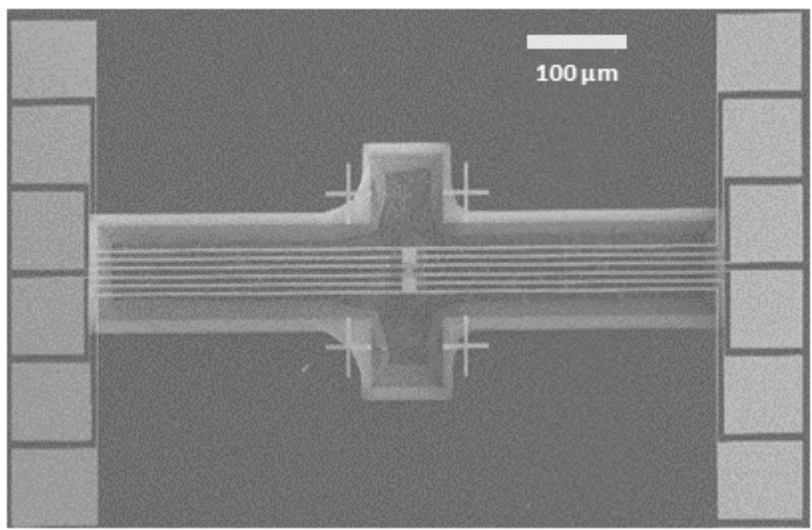

**Figure S1.** SEM image of a suspended microdevice showing the contact pads connected to suspended platforms at the center through suspended beams.

**Fabrication of thermal bridge device**

The fabrication of the suspended device involves several stages to accommodate the different feature sizes present in the device (see Figure S2). The first stage utilizes a laser writer, while the second stage employs electron beam lithography for patterning the smallest features with high precision. Subsequently, steps of dry and wet etching are carried out to suspend the microstructures.

**$SiN_x$ deposition** – The fabrication process begins by depositing a 525 nm thick layer of $SiN_x$ onto a two-inch undoped <110> silicon wafer. The deposition is performed using a plasma-enhanced chemical vapor deposition (PECVD) technique with a PlasmaPro PP80 PECVD system by Oxford instruments, which yield good-quality films [2]. During the deposition process, achieving a high-quality film with low strain is crucial for the successful suspension of the device at the final stage. To accomplish this, high and low frequency plasma conditions are applied for 16 and 4 second cycles, respectively, at supplied power of 20 W to alternatively deposit film with compressive and tensile strain to achieve a $SiN_x$ film with very low stress [3].

**Laser lithography** – After the $SiN_x$ deposition, the next step in the fabrication process involves the creation of the contact pads and lines using optical lithography. The wafer is cleaned with acetone and IPA followed by prebaking for 5 minutes. To achieve a clean lift-off process, two different positive resists are utilized to form a bi-layer to achieve an undercut profile. First, LOR3A resist is spin-coated onto the wafer at 4000 rpm for 45 seconds, followed by a soft baking step at 190 ºC for 2 minutes. Next, S1805 resist is spin-coated at 4500 rpm for 40 seconds and soft baked at 125 ºC for 2 minutes. Then the wafer is aligned in the laser writer, and the desired patterns are exposed onto the resist layer. Subsequently, the wafer is developed using a 2% tetramethylammonium hydroxide (TMAH) MF-319 solution for 1 minute, which selectively removes the exposed resist. To ensure complete removal of the developing solution and residues, the wafer is rinsed several times in deionized (DI) water.

**Metal deposition (Au)** – To create the metal contact pads and lines, a titanium (Ti) layer with a thickness of 3 nm, followed by a thin layer of gold (Au) with a thickness of 27 nm, is deposited using Electron Beam Evaporation. Before the deposition process, the sample undergoes a cleaning step using oxygen plasma for 10 seconds to remove any residual resist from the exposed regions. The deposition of Ti serves as an adhesion layer to the $SiN_x$ substrate. For lift-off, the sample is immersed in acetone in a bath set at a temperature of 50 ºC for a duration of 1 hour.

**Electron beam lithography** – First the wafer is cleaned by Oxygen plasma (60W) for 5 minutes to ensure that no resist residual is left after the lift off. In this case, PMMA is spin coated (5000 rpm, 40s) and baked at 180 ºC for 5 minutes. After the exposure, the wafer is developed in solution of 2% MIBK and IPA in a ratio of 1:3 for 1 minute 10 seconds and rinsed in IPA.

**Metal deposition (Pt)** – The resistors are fabricated by depositing a 3 nm layer of Titanium (Ti) followed by a 27 nm layer of Platinum (Pt) using Electron Beam Evaporation. Again, Ti acts as an adhesion layer. The lift-off process is performed in acetone at 50 ºC to for 1 hour followed by rinsing in IPA.

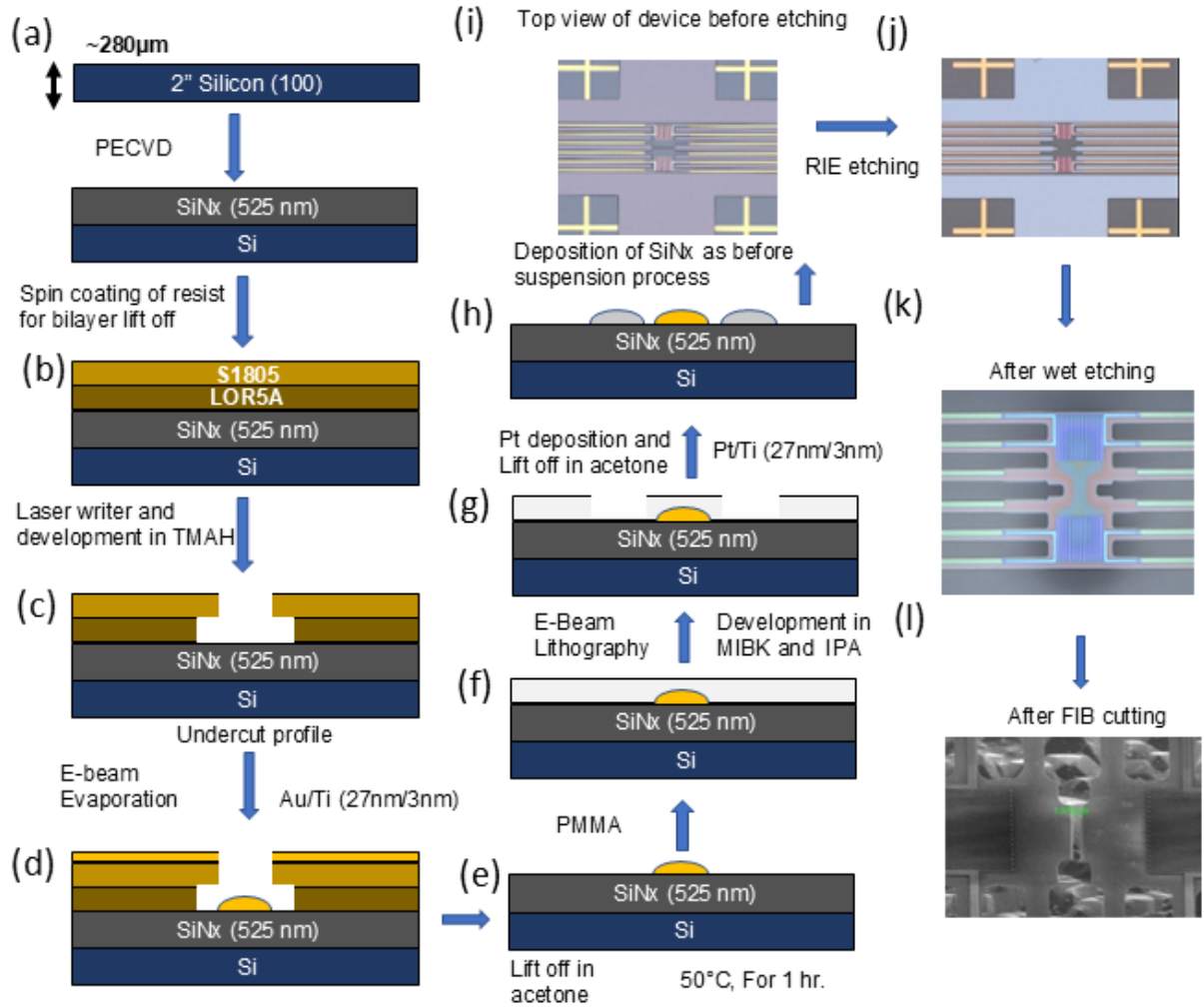

**Figure S2.** Schematic of fabrication steps for a suspended microdevice. a) CVD $SiN_x$ layer deposition. b-e) Lift-off process for patterning the Au contacts and beams using laser lithography. f-h) Lift-off process for patterning the Pt contacts and beams using e-beam lithography. i) Top view of the device after SiNx deposition to protect the metals during etching process. j) RIE etching of the $SiN_x$. k) Wet etching of the devices to suspend them. l) FIB cutting of the micro-platforms to pattern the desired gap.

**Device Suspension** – To enable proper thermal isolation and create a temperature gradient across the resistors, the device undergoes a suspension process. The regions surrounding the metal lines are selectively etched to remove the substrate material, to ensure that heat generated in the resistors is not dissipated into the substrate. Meanwhile, resist masks are used to protect the metal lines during the etching process. This suspension step is crucial for enabling effective heat transport through the NW and accurate thermal conductivity measurements.

To protect the metal lines during the wet etching process, a 450 nm layer of $SiN_x$ is deposited on the entire wafer using the PECVD. A photoresist layer (S1805) is then spin-coated onto the $SiN_x$ layer and patterned using optical lithography to create an opening in the mask for the etching region. The exposed sample is developed in a 2% TMAH (MF-319) solution for 1 minute. Reactive ion etching is used to selectively etch away the $SiN_x$ layer underneath the exposed regions, utilizing a mixture of $CHF_3$ and $O_2$ plasma. The process is carried out at 60 W applied power, 60 mTorr pressure, and gas flows of 80/8 sccm for $CHF_3/O_2$, resulting in an etch rate of approximately 35 nm per minute. The thickness of the photoresist is chosen such that it is etched away at the similar time as the $SiN_x$, ensuring complete removal of the resist and $SiN_x$ from the exposed areas. For silicon etching, a wet etching process is carried out using TMAH at 80 ºC for 7 hours, followed by KOH etching at 50 ºC for 10 minutes [4]. This etching process removes the silicon substrate in the areas where the $SiN_x$ layer has been etched away by dry etching process, suspending the beams and platforms with metal resistors and lines.

**Focused Ion Beam** – In the final step of the fabrication process, the two platforms of the device are separated using a focused ion beam (FIB). Using the FIB, a desirable gap of approximately 1 µm is created between the two suspended platforms.

## S3: Device characterization

Before measuring the thermal conductivity of the NWs, it is crucial to determine the thermal conductance of the suspended beams ($G_B$). The beam conductance $G_B$ represents the thermal conductance from the suspended membrane to the surrounding environment at a reference temperature $T_0$. This conductance accounts for the heat transfer occurring between the suspended membrane and the environment. The thermal conductance $G_B$ is calculated using Fourier's law (Equation 1.1). $\dot{Q}$ is the rate of heat transfer and $\Delta T$ is the temperature difference.

$$\dot{Q} = G_B \cdot \Delta T \tag{1.1}$$

The heat is generated through Joule heating in the platinum resistors. The power P dissipated in the resistors equals the heat transfer rate, $\dot{Q} = P$, and equation 1.1 can written as

$$P = G_B \cdot \Delta T \tag{1.2}$$

with $P = R \cdot I^2$, where P represents the power, R is the resistance of the resistors, and I is the current flowing through the resistors and depicted in Figure S3 (a). Assuming a linear relationship between the power dissipated in the resistors and the temperature rise in the suspended membrane, Equation 1.2 can be written as

$$G_B = \frac{P}{\Delta T} = \frac{dP}{dT} = \frac{dR/dT}{dR/dP}. \tag{1.3}$$

The power dissipated in the resistor leads to a temperature rise in the suspended membrane resulting in a temperature difference, $\Delta T$, between the center platform and the ends of the beams (contact pads). By measuring the resistance change in the resistors, where the majority of power is dissipated (platinum resistance is much higher than gold lines), the temperature rise in the suspended membrane can be determined. Figure 3 (a) shows the schematic of the measurement system for the one platform of the suspended device. Where $T_0$ is temperature of the substrate, T is temperature of the center platform, P is the power dissipated in the Pt resistor. The current is applied using multi probe tips and the voltage is measured to calculate the resistance of the Pt resistors.

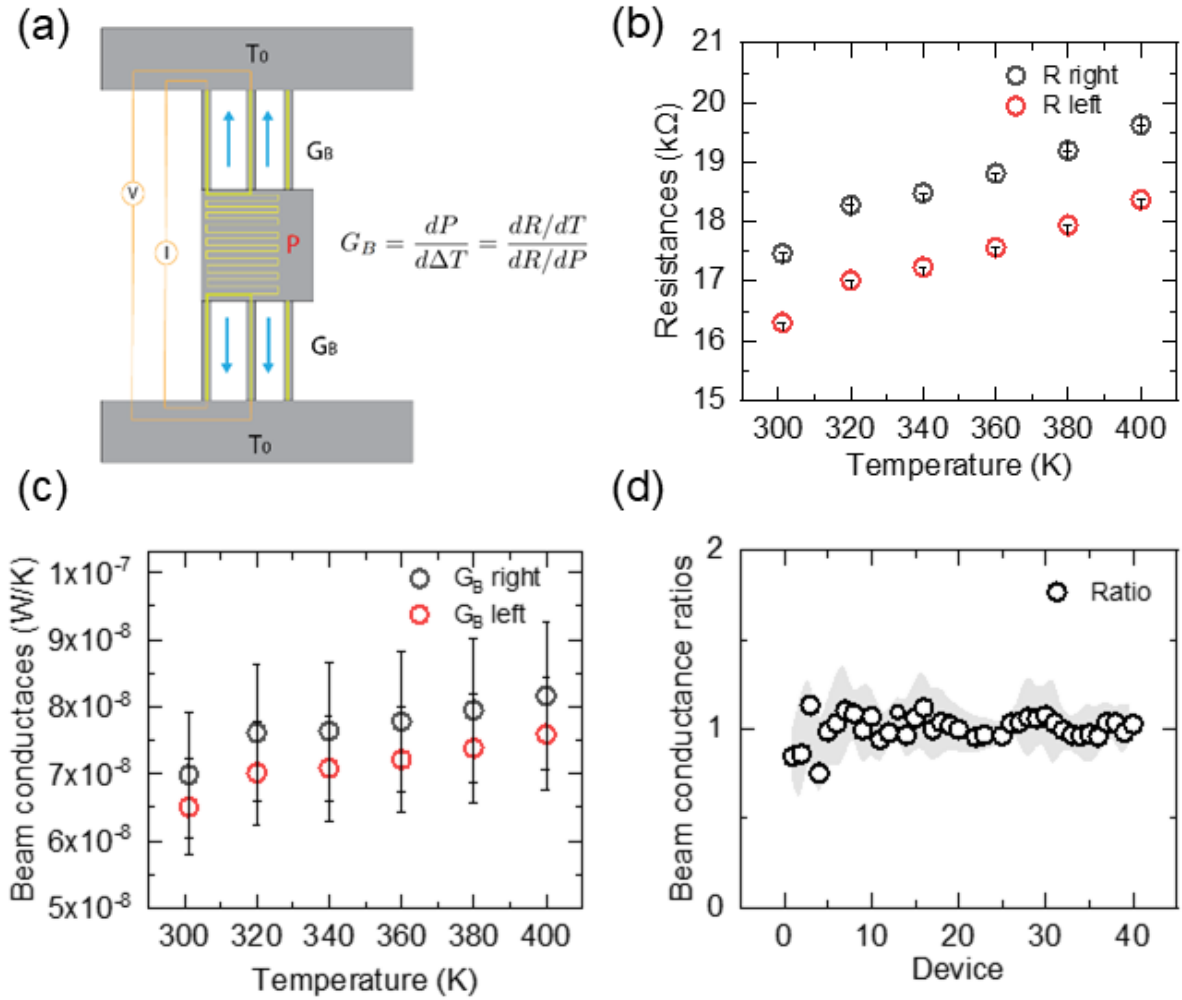

**Figure S3.** (a) Schematic showing the method for beam conductance measurements. (b) Resistance of platinum resistors with respect to temperature. (c) Beam conductance of left and right beams with temperature, and (d) beam conductance ratio of all the measured devices showing the device asymmetry.

**Measurement Steps for Device Characterization:**

Temperature Stabilization: The sample stage is first heated to a specific base temperature to ensure a stable and controlled thermal environment. This step allows the device to reach thermal equilibrium before measurements begin, minimizing any transient thermal effects.

Resistance Measurements: Small currents, ranging from -450 nA to 450 nA, are applied to the device. At each current level, the 4-point resistance is measured, ensuring precise readings by eliminating contact resistance and the Au line contributions. The measured resistance values are then plotted as a function of temperature. From this data, the temperature derivative of resistance, dR/dT, is calculated (Figure S3 (b)).

Gradient Application: To extract the derivative dR/dP, a gradient is applied to the device. This additional measurement step helps quantify how resistance varies with an increasing power.

Using the obtained values of dR/dT and dR/dP, the beam conductance $G_B$ is calculated based on Equation 1.3. Figure S3 (c) illustrates the measured conductance for both the left and right beams as a function of temperature, beam conductance increases with rising temperature, but the ratio between the left and right beam conductance remains constant. Additionally, Figure S3 (d) presents the beam conductance ratio for all measured devices, with the x-axis representing the device number. This illustrates the variation in the asymmetry factor (the conductance ratio of left and right beam) of the fabricated devices[5].

### S4: Nanowire conductance measurements

To assess the thermal conductance of the NW, we start by passing current through the beams to one of the resistors, inducing a temperature increase $\Delta T_H$ on the heater side of the platform due to

Joule heating. As heat passes through the NW, a small $\Delta T_S$ is observed on the sensor platform. The schematic in Figure S4 depicts the components of the thermal bridge and their respective heat fluxes. We estimate the heat flow through the NW by measuring the power dissipated in the resistors via a 4-point measurement.

Once we have determined the total power heating up the platforms, we can apply the principle of energy conservation to the entire system to estimate the conductance of the beams as follows:

$$P_H + P_S = G_{B,H}\Delta T_H + G_{B,S}\Delta T_S \qquad (2.1)$$

where $P_H$ and $P_S$ represent the power dissipated in the heater and sensor resistors, respectively; $G_{B,H}$ and $G_{B,S}$ denote the beam conductance on the heater and sensor sides, respectively; $\Delta T_H$ and $\Delta T_S$ represent the temperature rise with respect to the base temperature $T_0$ on the heater and sensor sides, respectively.

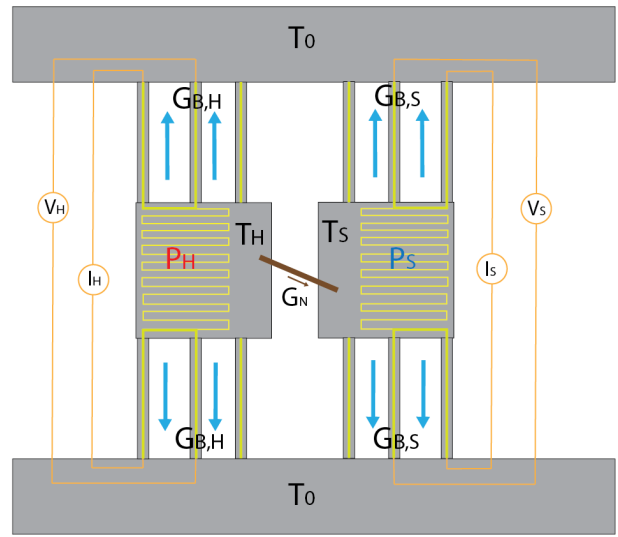

**Figure S4.** Schematic illustration of suspended microdevice used for measuring the thermal conductivity of NWs. The direction of heat flow is indicated by arrows.

From the energy balance on the sensor side, we obtain:

$$G_N(\Delta T_H - \Delta T_S) + P_S = G_{B,S}\Delta T_S . \tag{2.2}$$

From equation (2.1):

$$G_{B,S} = (P_H + P_S) \cdot \left(\frac{G_{B,H}}{G_{B,S}}\Delta T_H + \Delta T_S\right)^{-1} . \tag{2.3}$$

In order to use slopes from the measurements of power and temperature, we assume that power and temperature follow a linear relationship. Thus, we can write

$$\frac{d\Delta T_x}{dP} = \Delta T_x / (P_H + P_S) \tag{2.4}$$

and rewrite Equation (2.3) as:

$$G_{B,S} = \left(\frac{G_{B,H}}{G_{B,S}} \cdot \frac{d\Delta T_H}{dP} + \frac{d\Delta T_S}{dP}\right)^{-1} . \tag{2.5}$$

Using the Equation 2.2, the NW conductance $G_N$ can be given as

$$G_N = G_{B,S}\frac{d\Delta T_S}{dP}\left(\frac{G_{B,H}}{G_{B,S}} \cdot \frac{d\Delta T_H}{dP} - \frac{d\Delta T_S}{dP}\right)^{-1} \tag{2.6}$$

where $\frac{d\Delta T_H}{dP}$ and $\frac{d\Delta T_S}{dP}$ are the slopes of temperature rise on the heater and sensor side, respectively, as a function of total power dissipated, and $G_{B,H}$ and $G_{B,S}$ are the heater and sensor beam conductance, respectively. $\frac{G_{B,H}}{G_{B,S}}$ is defined the asymmetry factor of the device beams' conductance.

The thermal conductivity of the measured NW can be extracted using the following Equation

$$k = \frac{4G_N L}{\pi D^2}$$

where L and D are the suspended length and diameter of the NW [1,6].

## S5: Details of the theoretical calculations

**Lattice dynamics calculations.** We perform density-functional theory (DFT) calculations with the VASP [7,8] code, projector-augmented wave (PAW) potentials [9,10], and the local density approximation (LDA) for the exchange correlation energy. The unit cells of bulk wurtzite GaAs and GaP were optimized until strict convergence criteria for stress ($3\times10^{-3}$ GPa) and forces ($5\times10^{-4}$ eV/Å), sampling the Brillouin zone with a $16\times16\times12$ $\Gamma$-centered mesh. Harmonic and third-order anharmonic interatomic force constants (IFCs), were computed in $4\times4\times3$ supercells using the phonopy [11] and the thirdorder.py code [12]. Interactions beyond fourth nearest neighbors were neglected in the case of anharmonic IFCs. The linearized phonon Boltzmann Transport Equation (BTE) was solved with the methods implemented in the almaBTE code [13], where superlattices are modelled as a periodic perturbation upon a reference virtual crystal [14]. We consider rather sharp, but gradual interfaces, where the chemical composition switch from GaAs to GaP within two-unit cells (the composition if 100% GaAs in cell $i$, 75% GaAs and 25% GaP in cell $i+1$, 25% GaAs and 75% GaP in cell $i+2$, and finally 100% GaP in cell $i+3$). We also considered idealized, atomically flat interfaces, where phonon scattering is entirely dominated by Umklapp processes, due to the low value of the thermal boundary resistance (TBR). In this case $\kappa(L)$ flattens, rather than increasing, as L increases (see Figure S5), because of the low contribution of the TBR of each interface (having e.g. 4 or 5 interfaces results in a negligible difference).

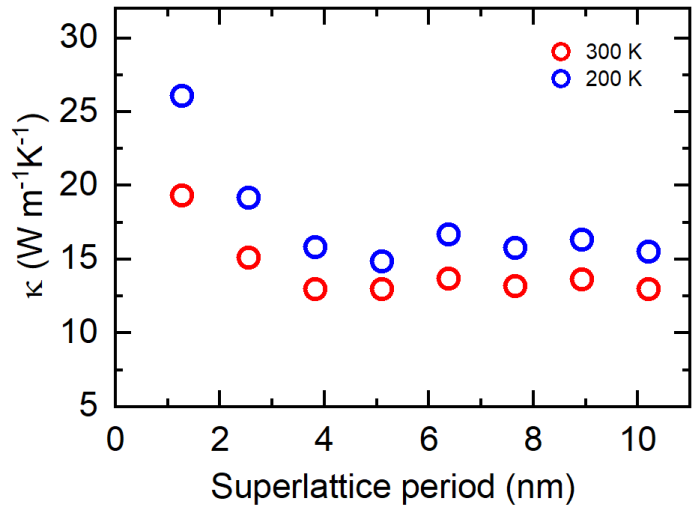

**Figure S5.** Computed thermal conductivity as function of the period in GaAs/GaP SLs with idealized flat interfaces.

In Figure S6 we plot the dependence of the thermal conductivity as a function of temperature, κ(T), for all the SL investigated. As it can be seen, for the shortest period, when transport is fully coherent, κ(T) resembles that of a homogeneous material, with a rater sharp peak at low temperature. As the period increases and coherence is progressively lost, the peak shifts toward higher temperature and becomes blunter. This trend continues until the period reaches 5.1 nm, which is where we observe the coherent-incoherent crossover and is then reversed, with the maximum of κ(T) shifting back to lower temperature and tending to become sharper (this trend would become more evident for longer periods, which we cannot afford computationally). Notice that for temperatures ≲ 50 K these results have to be taken only qualitatively, because transport is dominated by phonons of increasingly long wavelength and the **q-point** meshes required to achieve strict convergence criteria cannot be handled efficiently.

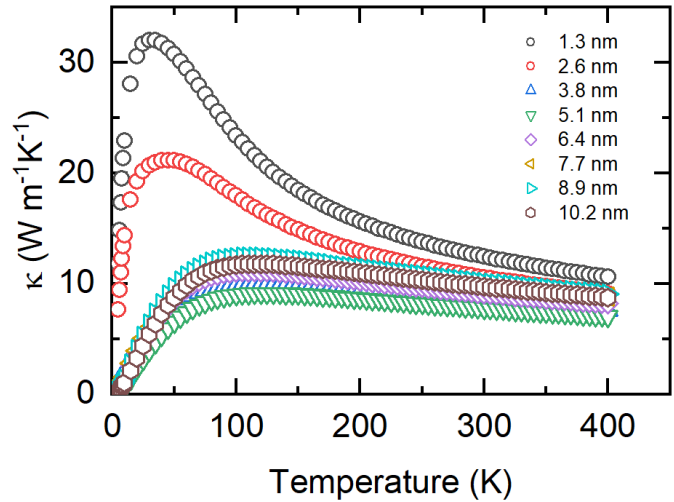

**Figure S6.** Computed thermal conductivity as function of temperature in GaAs/GaP SLs of different periods, L.

**Molecular dynamics calculations.** We performed nonequilibrium molecular dynamics (NEMD) simulations with the LAMMPS code [15,16] by imposing a temperature bias of 170 K, with the hot and cold reservoirs set at 385 K and 215 K, respectively. The timestep was set to 0.7 fs, and the total simulation time was 7 ns. After an initial structural relaxation, the temperatures of the simulation cell ends are controlled by rescaling the velocities of the atoms therein, while the central region is left free to evolve without constraints; full details of the simulation protocol can be found e.g. in Ref. [17]. We account for the atomic interactions by mean of the bond-order classical potential due to Powell *et al* [18]. The mixed interactions between atoms belonging to neighboring GaP and GaAs segments were derived using the empirical approach of Tersoff in the case of SiC and SiGe [19]. The NW structure was generated by stacking the appropriate supercell until reaching the target length of approximately 100 nm. Subsequently, atoms outside the hexagonal cross-section with an edge length of 2.5 nm were removed, resulting in a NW with a final diameter of 5 nm. The NWs studies are oriented along the <111> crystallographic axis and have a length of approximately 100 nm.